\documentclass[aps,prl,twocolumn,superscriptaddress]{revtex4-2}

\usepackage{graphicx}
\usepackage{dcolumn}
\usepackage{bm}
\usepackage{subfigure}
\usepackage{mathrsfs}
\usepackage{multirow}
\usepackage{amsmath}
\usepackage{orcidlink}
\usepackage{hyperref}
\usepackage{xspace}
\usepackage{amssymb}
\usepackage{braket}
\usepackage{bm}  
\usepackage[capitalise]{cleveref}

\usepackage{tikz}
\usepackage[compat=1.1.0]{tikz-feynman}
\usepackage{booktabs}
\tikzset{every picture/.style={line width=0.9pt}}

\newcommand{\Cdiagram}[3]{
\begin{tikzpicture}
\begin{feynman}
  \def\dx{0.4}   
  \def\dy{0.7}   
  \def\labelscale{1.3} 

  \vertex (c) at (0,0);

  \vertex (f1) at (-\dx, +\dy);
  \vertex (f2) at ( +\dx, +\dy);
  \vertex (i1) at (-\dx, -\dy);
  \vertex (i2) at ( +\dx, -\dy);

  \diagram*{
    (i1) -- [plain] (c) -- [plain] (i2),
    (f1) -- [plain] (c) -- [plain] (f2),
  };

  \node[draw, #1, fill=#2, minimum size=#3, inner sep=0pt, color=#2] at (c) {};

\end{feynman}
\end{tikzpicture}
}

\newcommand{\Ctriangle}{
\begin{tikzpicture}
\begin{feynman}
  \def\dx{0.4}   
  \def\dy{0.7}   
  \def\labelscale{1.3} 

  \vertex (c) at (0,0);

  \vertex (f1) at (-\dx, +\dy);
  \vertex (f2) at ( +\dx, +\dy);
  \vertex (i1) at (-\dx, -\dy);
  \vertex (i2) at ( +\dx, -\dy);

  \diagram*{
    (i1) -- [plain] (c) -- [plain] (i2),
    (f1) -- [plain] (c) -- [plain] (f2),
  };

  \node[draw, isosceles triangle, fill=magenta, minimum size=5pt,inner sep=0pt, rotate=90, color=magenta] at (c) {};

\end{feynman}
\end{tikzpicture}
}

\newcommand{\CLOdiagram}{
\begin{tikzpicture}
\begin{feynman}
  \def\dx{0.4}   
  \def\dy{0.7}   
  \def\labelscale{1.3} 

  \vertex (c) at (0,0);

  \vertex (f1) at (-\dx, +\dy);
  \vertex (f2) at ( +\dx, +\dy);
  \vertex (i1) at (-\dx, -\dy);
  \vertex (i2) at ( +\dx, -\dy);

  \diagram*{
    (i1) -- [plain] (c) -- [plain] (i2),
    (f1) -- [plain] (c) -- [plain] (f2),
  };

  \node[draw, fill=black, circle, inner sep=1pt] at (c) {};

\end{feynman}
\end{tikzpicture}
}

\newcommand{\OPEdiagram}{
\begin{tikzpicture}
\begin{feynman}
  \def\dx{0.4}   
  \def\dy{0.7}   
  \def\labelscale{1.3} 

  \vertex (c) at (0,0);

  \vertex (f1) at (-\dx, +\dy);
  \vertex (f2) at ( +\dx, +\dy);
  \vertex (i1) at (-\dx, -\dy);
  \vertex (i2) at ( +\dx, -\dy);
  \vertex (c1) at (-\dx, 0);
  \vertex (c2) at ( +\dx, 0);
  \diagram*{
    (i1) -- [plain] (f1),
    (c1) -- [scalar] (c2),
    (i2) -- [plain] (f2),
  };

  \node[draw, fill=black, circle, inner sep=1pt] at (c1) {};
  \node[draw, fill=black, circle, inner sep=1pt] at (c2) {};
 
\end{feynman}
\end{tikzpicture}
}

\newcommand{\TPEbox}{
\begin{tikzpicture}
\begin{feynman}
  \def\dx{0.4}   
  \def\dy{0.7}   
  \def\labelscale{1.3} 

  \vertex (c) at (0,0);

  \vertex (f1) at (-\dx, +\dy);
  \vertex (f2) at ( +\dx, +\dy);
  \vertex (i1) at (-\dx, -\dy);
  \vertex (i2) at ( +\dx, -\dy);
  \vertex (cl1) at (-\dx, -\dy/1.75);
  \vertex (ch2) at ( +\dx, +\dy/1.75);
  \vertex (ch1) at (-\dx, +\dy/1.75);
  \vertex (cl2) at ( +\dx, -\dy/1.75);
  \diagram*{
    (i1) -- [plain] (f1),
    (ch1) -- [scalar] (ch2),
    (cl1) -- [scalar] (cl2),
    (i2) -- [plain] (f2),
  };

  \node[draw, fill=black, circle, inner sep=1pt] at (cl1) {};
  \node[draw, fill=black, circle, inner sep=1pt] at (cl2) {};
  \node[draw, fill=black, circle, inner sep=1pt] at (ch1) {};
  \node[draw, fill=black, circle, inner sep=1pt] at (ch2) {};

\end{feynman}
\end{tikzpicture}
}

\newcommand{\TPEtrianglel}{
\begin{tikzpicture}
\begin{feynman}
  \def\dx{0.4}   
  \def\dy{0.7}   
  \def\labelscale{1.3} 

  \vertex (c) at (0,0);

  \vertex (f1) at (-\dx, +\dy);
  \vertex (f2) at ( +\dx, +\dy);
  \vertex (i1) at (-\dx, -\dy);
  \vertex (i2) at ( +\dx, -\dy);
  \vertex (c1) at (-\dx, 0);
  \vertex (c2) at ( +\dx, 0);
  \vertex (cl1) at (-\dx, -\dy/1.75);
  \vertex (ch2) at ( +\dx, +\dy/1.75);
  \vertex (ch1) at (-\dx, +\dy/1.75);
  \vertex (cl2) at ( +\dx, -\dy/1.75);
  \diagram*{
    (i1) -- [plain] (f1),
    (ch1) -- [scalar] (c2),
    (cl1) -- [scalar] (c2),
    (i2) -- [plain] (f2),
  };

  \node[draw, fill=black, circle, inner sep=1pt] at (cl1) {};
  \node[draw, fill=black, circle, inner sep=1pt] at (ch1) {};
  \node[draw, fill=black, circle, inner sep=1pt] at (c2) {};

\end{feynman}
\end{tikzpicture}
}

\newcommand{\TPEtrianglell}{
\begin{tikzpicture}
\begin{feynman}
  \def\dx{0.4}   
  \def\dy{0.7}   
  \def\labelscale{1.3} 

  \vertex (c) at (0,0);

  \vertex (f1) at (-\dx, +\dy);
  \vertex (f2) at ( +\dx, +\dy);
  \vertex (i1) at (-\dx, -\dy);
  \vertex (i2) at ( +\dx, -\dy);
  \vertex (c1) at (-\dx, 0);
  \vertex (c2) at ( +\dx, 0);
  \vertex (cl1) at (-\dx, -\dy/1.75);
  \vertex (ch2) at ( +\dx, +\dy/1.75);
  \vertex (ch1) at (-\dx, +\dy/1.75);
  \vertex (cl2) at ( +\dx, -\dy/1.75);
  \diagram*{
    (i1) -- [plain] (f1),
    (ch1) -- [scalar] (c2),
    (cl1) -- [scalar] (c2),
    (i2) -- [plain] (f2),
  };

  \node[draw, fill=black, circle, inner sep=1pt] at (cl1) {};
  \node[draw, fill=black, circle, inner sep=1pt] at (ch1) {};
  \node[draw, fill=blue, circle, inner sep=1.6pt,color=blue] at (c2) {};

\end{feynman}
\end{tikzpicture}
}

\newcommand{\TPEtriangler}{
\begin{tikzpicture}
\begin{feynman}
  \def\dx{0.4}   
  \def\dy{0.7}   
  \def\labelscale{1.3} 

  \vertex (c) at (0,0);

  \vertex (f1) at (-\dx, +\dy);
  \vertex (f2) at ( +\dx, +\dy);
  \vertex (i1) at (-\dx, -\dy);
  \vertex (i2) at ( +\dx, -\dy);
  \vertex (c1) at (-\dx, 0);
  \vertex (c2) at ( +\dx, 0);
  \vertex (cl1) at (-\dx, -\dy/1.75);
  \vertex (ch2) at ( +\dx, +\dy/1.75);
  \vertex (ch1) at (-\dx, +\dy/1.75);
  \vertex (cl2) at ( +\dx, -\dy/1.75);
  \diagram*{
    (i1) -- [plain] (f1),
    (c1) -- [scalar] (cl2),
    (c1) -- [scalar] (ch2),
    (i2) -- [plain] (f2),
  };

  \node[draw, fill=black, circle, inner sep=1pt] at (cl2) {};
  \node[draw, fill=black, circle, inner sep=1pt] at (ch2) {};
  \node[draw, fill=black, circle, inner sep=1pt] at (c1) {};

\end{feynman}
\end{tikzpicture}
}

\newcommand{\TPEtrianglerr}{
\begin{tikzpicture}
\begin{feynman}
  \def\dx{0.4}   
  \def\dy{0.7}   
  \def\labelscale{1.3} 

  \vertex (c) at (0,0);

  \vertex (f1) at (-\dx, +\dy);
  \vertex (f2) at ( +\dx, +\dy);
  \vertex (i1) at (-\dx, -\dy);
  \vertex (i2) at ( +\dx, -\dy);
  \vertex (c1) at (-\dx, 0);
  \vertex (c2) at ( +\dx, 0);
  \vertex (cl1) at (-\dx, -\dy/1.75);
  \vertex (ch2) at ( +\dx, +\dy/1.75);
  \vertex (ch1) at (-\dx, +\dy/1.75);
  \vertex (cl2) at ( +\dx, -\dy/1.75);
  \diagram*{
    (i1) -- [plain] (f1),
    (c1) -- [scalar] (cl2),
    (c1) -- [scalar] (ch2),
    (i2) -- [plain] (f2),
  };

  \node[draw, fill=black, circle, inner sep=1pt] at (cl2) {};
  \node[draw, fill=black, circle, inner sep=1pt] at (ch2) {};
  \node[draw, fill=blue, circle, inner sep=1.6pt,color=blue] at (c1) {};

\end{feynman}
\end{tikzpicture}
}

\newcommand{\TPEloop}{
\begin{tikzpicture}
\begin{feynman}
  \def\dx{0.4}   
  \def\dy{0.7}   
  \def\labelscale{1.3} 

  \vertex (c) at (0,0);

  \vertex (f1) at (-\dx, +\dy);
  \vertex (f2) at ( +\dx, +\dy);
  \vertex (i1) at (-\dx, -\dy);
  \vertex (i2) at ( +\dx, -\dy);
  \vertex (c1) at (-\dx, 0);
  \vertex (c2) at ( +\dx, 0);
  \diagram*{
    (i1) -- [plain] (f1),
   
    (i2) -- [plain] (f2),

    (c1) -- [scalar, out=60, in=120, looseness=1.4] (c2),
    (c1) -- [scalar, out=-60, in=-120, looseness=1.4] (c2),
  };

  \node[draw, fill=black, circle, inner sep=1pt] at (c1) {};
  \node[draw, fill=black, circle, inner sep=1pt] at (c2) {};

\end{feynman}
\end{tikzpicture}
}

\newcommand{\TPEloopl}{
\begin{tikzpicture}
\begin{feynman}
  \def\dx{0.4}   
  \def\dy{0.7}   
  \def\labelscale{1.3} 

  \vertex (c) at (0,0);

  \vertex (f1) at (-\dx, +\dy);
  \vertex (f2) at ( +\dx, +\dy);
  \vertex (i1) at (-\dx, -\dy);
  \vertex (i2) at ( +\dx, -\dy);
  \vertex (c1) at (-\dx, 0);
  \vertex (c2) at ( +\dx, 0);
  \diagram*{
    (i1) -- [plain] (f1),
   
    (i2) -- [plain] (f2),

    (c1) -- [scalar, out=60, in=120, looseness=1.4] (c2),
    (c1) -- [scalar, out=-60, in=-120, looseness=1.4] (c2),
  };

  \node[draw, fill=black, circle, inner sep=1pt] at (c1) {};
  \node[draw, fill=blue, circle, inner sep=1.6pt,color=blue] at (c2) {};

\end{feynman}
\end{tikzpicture}
}

\newcommand{\TPEloopr}{
\begin{tikzpicture}
\begin{feynman}
  \def\dx{0.4}   
  \def\dy{0.7}   
  \def\labelscale{1.3} 

  \vertex (c) at (0,0);

  \vertex (f1) at (-\dx, +\dy);
  \vertex (f2) at ( +\dx, +\dy);
  \vertex (i1) at (-\dx, -\dy);
  \vertex (i2) at ( +\dx, -\dy);
  \vertex (c1) at (-\dx, 0);
  \vertex (c2) at ( +\dx, 0);
  \diagram*{
    (i1) -- [plain] (f1),
   
    (i2) -- [plain] (f2),

    (c1) -- [scalar, out=60, in=120, looseness=1.4] (c2),
    (c1) -- [scalar, out=-60, in=-120, looseness=1.4] (c2),
  };

  \node[draw, fill=blue, circle, inner sep=1.6pt,color=blue] at (c1) {};
  \node[draw, fill=black, circle, inner sep=1pt] at (c2) {};

\end{feynman}
\end{tikzpicture}
}

\newcommand{\TPEcrossedbox}{
\begin{tikzpicture}
\begin{feynman}
  \def\dx{0.4}   
  \def\dy{0.7}   
  \def\labelscale{1.3} 

  \vertex (c) at (0,0);

  \vertex (f1) at (-\dx, +\dy);
  \vertex (f2) at ( +\dx, +\dy);
  \vertex (i1) at (-\dx, -\dy);
  \vertex (i2) at ( +\dx, -\dy);
  \vertex (cl1) at (-\dx, -\dy/1.75);
  \vertex (ch2) at ( +\dx, +\dy/1.75);
  \vertex (ch1) at (-\dx, +\dy/1.75);
  \vertex (cl2) at ( +\dx, -\dy/1.75);
  \diagram*{
    (i1) -- [plain] (f1),
    (ch1) -- [scalar] (cl2),
    (cl1) -- [scalar] (ch2),
    (i2) -- [plain] (f2),
  };

        \node[draw, fill=black, circle, inner sep=1pt] at (cl1) {};
  \node[draw, fill=black, circle, inner sep=1pt] at (cl2) {};
  \node[draw, fill=black, circle, inner sep=1pt] at (ch1) {};
  \node[draw, fill=black, circle, inner sep=1pt] at (ch2) {};

\end{feynman}
\end{tikzpicture}
}

\newcommand{\chEFT}{$\chi$EFT}

\newcommand*{\LO}{LO}
\newcommand*{\NLO}{NLO}
\newcommand*{\NNLO}{N$^2$LO}
\newcommand*{\NNNLO}{N$^3$LO}

\renewcommand*{\H}{$^3$H}
\newcommand*{\He}{$^4$He}
\newcommand*{\Li}{$^6$Li}
\newcommand*{\Lambdabreak}{\Lambda_b}

\newcommand{\Nmax}{N_\mathrm{max}}

\newcommand*{\tS}{{}^3S_1 \mathrm{-} {}^3D_1}
\newcommand*{\tP}{{}^3P_2 \mathrm{-} {}^3F_2}

\newcommand*{\Tl}{T_\mathrm{lab}}

\begin{document}

\title{Perturbative calculations of light nuclei up to N$^3$LO in chiral effective field theory}

\author{Oliver Thim} \email{oliver.thim@chalmers.se}
\author{Andreas Ekström}
\author{Christian Forssén}
\affiliation{Department of Physics and Astronomy, Chalmers University of Technology, SE-412 96, Göteborg, Sweden}

\begin{abstract}
We predict ground-state energies of $^3$H, $^4$He, and $^6$Li in chiral effective field theory up to next-to-next-to-next-to-leading-order (N$^3$LO) using a power counting guided by renormalization-group invariance. Subleading two-nucleon interactions are treated perturbatively, and for $^4$He and $^6$Li, we calculate the perturbative corrections from numerical derivatives of ground-state energies obtained with Lanczos diagonalization. We find that including the $^3$H binding energy in the calibration is essential for robust predictions of $^4$He and $^6$Li. This work demonstrates that the employed power counting can be applied to construct nuclear interactions with predictive power for light nuclei, bringing nuclear structure predictions closer to a foundation in quantum chromodynamics.
\end{abstract}

\maketitle

{\it Introduction.$-$}
A central challenge in nuclear effective field theory is to construct nuclear interactions, guided by renormalization-group (RG) invariance, and accurately predict the properties of atomic nuclei. Chiral effective field theory (\chEFT) \cite{Weinberg:1978kz, Weinberg:1990rz,Weinberg:1991um} provides a framework to analyze strongly interacting nucleons consistently with the symmetries of low-energy quantum chromodynamics (QCD)~\cite{Machleidt:2011zz,Epelbaum:2008ga,Hammer:2019poc}. A power counting (PC) scheme orders interaction terms according to their expected contributions to observables. This allows a systematic truncation of the sum of interaction diagrams derived from the effective Lagrangian. 
The vast majority of \chEFT‑based \textit{ab initio} nuclear calculations~\cite{Hergert:2020bxy,Ekstrom:2022yea} employ Weinberg power counting (WPC)~\cite{Weinberg:1990rz,Weinberg:1991um}.

While WPC has proven successful in reproducing and predicting selected nuclear observables~\cite{Elhatisari:2015iga,Stroberg:2019bch,Hu:2021trw,Sun:2024iht}, it violates RG invariance at the nucleon-nucleon (NN) level~\cite{Nogga:2005hy} due to uncontrolled short-distance behavior of pion-exchange potentials \cite{Frank:1971xx}. This motivates the exploration of modified PC schemes constructed to satisfy RG invariance, in which subleading orders are treated perturbatively~\cite{vanKolck:2020llt,Long:2007vp}. The manifest order-by-order hierarchy of a perturbative PC also provides a principled basis for estimating the EFT truncation error. However, the extent to which such PC schemes can achieve the predictive power required for nuclear structure and reaction studies remains largely untested.

Previous studies employing RG invariant \chEFT{} have investigated NN scattering up to \NNNLO{} \cite{Long:2012ve,PhysRevC.85.034002,Long:2011qx,Long:2007vp,PavonValderrama:2025azr,Valderrama:2009ei,Thim:2024yks,Thim:2024jdv} and $^3$H, $^{3,4}$He, $^6$Li, and $^{16}$O up to next-to-leading order (NLO) \cite{Song:2016ale,Yang:2020pgi}. Recently, we performed a study of \H{} at next-to-next-to-leading-order (\NNLO) \cite{Thim:2025vhe}, where so-called exceptional cutoffs \cite{Gasparyan:2022isg,Peng:2024aiz,Yang:2024yqv}---previously only studied in the NN system---challenged the RG invariance of this modified PC. Essential next steps for exploring whether modified PCs can provide realistic descriptions of nuclear observables are: $(i)$ the construction of interactions to a non-trivial chiral order, and $(ii)$ the development of computational frameworks for perturbative computations of nuclear observables beyond the NN system. 

In this work, we address these challenges by employing the modified PC proposed by Long and Yang \cite{Long:2012ve,PhysRevC.85.034002,Long:2011qx,Long:2007vp} to construct chiral NN interactions up to next-to-next-to-next-to-leading order (\NNNLO) and compute ground-state energies of \H, \He, and \Li{} using the no-core shell model (NCSM). 
Converged NCSM results are obtained by using relatively low values of the momentum space cutoff $\Lambda \approx 500$~MeV in the potential. This also implies that exceptional cutoffs are not expected to have a large impact \cite{Thim:2025vhe}. For $^3$H, we employ the Jacobi-coordinate NCSM (J-NCSM) code \texttt{py-ncsm} \cite{py-ncsm}, in which we explicitly implement Rayleigh-Schrödinger perturbation theory. We compute the ground-state energies for \He{} and \Li{} using an M-scheme NCSM (M-NCSM) code \texttt{pANTOINE} \cite{Forssen:2017wei,Caurier:1999,Caurier:1998zw,Navrtil:2004qx} and calculate the perturbative corrections using numerical derivatives of non-perturbative computations. Higher-order contributions in WPC has also been treated in perturbation theory up to second order in quantum Monte Carlo methods~\cite{Lu:2021tab,Curry:2023mkm,Curry:2024gcz} for nuclear systems.

{\it Nucleon-nucleon potential up to \NNNLO.$-$}
In \chEFT{}, the nuclear potential is constructed from irreducible Feynman diagrams, including pion-exchange contributions and short-range contact interactions. The PC quantifies the importance of a given diagram in terms of the expansion parameter $(Q/\Lambdabreak)^\nu$, where the pion-mass $m_\pi$ represents the typical low-energy scale $Q\approx m_\pi$ and $\Lambdabreak$ is the \chEFT{} breakdown scale. Estimates of the breakdown scale in WPC range between $\Lambdabreak\approx 500-600$~MeV ~\cite{Melendez:2019izc,Millican:2024yuz}, and is possibly even lower, $\Lambdabreak\approx 300$~MeV, in the Long and Yang PC \cite{Thim:2024yks}. The violation of RG-invariance in WPC, manifested as a significant cutoff dependence in observables, is due to the singular and attractive \cite{Frank:1971xx,Nogga:2005hy} nature of the pion-exchange potentials, which generate divergences when treated non-perturbatively without sufficient counterterms. At LO, these divergences are caused by the one-pion exchange (OPE)  potential, and can be absorbed by promoting nucleon-contact interactions. 

We employ a version \cite{Thim:2024yks} of the Long and Yang PC~\cite{Long:2012ve,PhysRevC.85.034002,Long:2011qx} to construct NN potentials up to \NNNLO, i.e., $(Q/\Lambdabreak)^3$. Here, only LO is treated non-perturbatively while subleading corrections are included perturbatively. It has been shown that the OPE potential needs to be treated non-perturbatively only for the lowest relative NN angular momenta~\cite{Birse:2005um, PhysRevC.99.024003, Peng:2020nyz}. Thus, OPE is considered LO up to $P$-waves, i.e., for the NN channels: ${}^1S_0$, ${}^3P_0$, ${}^1P_1$, $^3P_1$, $\tS$ and $\tP$, always including the full coupled channels. The contributions to the NN potential in these channels are shown diagrammatically in the left column of \cref{tab:pot_diagrams}. The underlined diagrams are higher-order contacts that are promoted to absorb cutoff dependence. For the remaining channels, there is no LO contribution, and the NN force is treated entirely perturbatively, see the rightmost column of \cref{tab:pot_diagrams}. We follow Ref.~\cite{PhysRevC.99.024003} and suppress two-pion exchanges (TPEs) by the same chiral order as OPE. Note that there are no contact interactions in these channels up to \NNNLO.

Besides the diagrams shown in \cref{tab:pot_diagrams}, there are corrections to OPE and lower-order contacts at \NNLO{} and \NNNLO{} \cite{Epelbaum:2008ga}. We absorb such OPE corrections at LO by renormalizing the axial coupling to $g_A=1.29$. The subleading contact contributions that provide constant shifts in the LECs at and below the given order need to be included at their respective order in perturbative computations. In practice, this is done via perturbative corrections to the LECs that are introduced at subsequent orders \cite{Thim:2024yks}.
For example, in the ${}^1S_0$ channel, there is one LEC at LO, and the contact part is simply $C^{(0)}_{{}^1S_0}$. At NLO, the contact potential reads $C^{(1)}_{{}^1S_0} + D^{(0)}_{{}^1S_0}(p'^2+p^2)$, where $C^{(1)}_{{}^1S_0}$ denotes the correction to $C^{(0)}_{{}^1S_0}$ and $D^{(0)}_{{}^1S_0}(p'^2+p^2)$ is the promoted $\nu=2$ operator (shown in \cref{tab:pot_diagrams}). This pattern continues at subsequent orders, for all channels, where $p$ $(p')$ denotes the modulus of the ingoing (outgoing) relative momenta.

As our starting point, we construct NN potentials $\{V^{(\nu)}\}_{\nu=0}^3$ by calibrating the unknown values of the 33 LECs, shown in \cref{tab:LECs} in End Matter, to neutron-proton phase shifts following Ref.~\cite{Thim:2024yks}. In a subsequent step, the calibration is refined by matching to the experimental values of the $^{2,3}$H binding energies, where all details are given in the End Matter.
\begin{table}[h!]
\centering
\caption{Diagrammatic representation of the contributions to the NN force in the PC employed in this work. Black dots, blue circles, orange squares, red diamonds, and pink triangles denote vertices with interaction index $\Delta_i=0,1,2,4,6$, respectively. See, e.g., Refs.~\cite{Machleidt:2011zz,Epelbaum:2008ga} for details. Underlined contact diagrams are promoted compared to WPC. All contributions beyond LO are treated perturbatively.}
{\renewcommand{\arraystretch}{0}
\begin{tabular}{c | c | c}
\toprule
\ \ Order \ \ & Channels: ${}^1S_0$, ${}^3P_0$, ${}^1P_1$,  & Remaining channels \\
 & $^3P_1$, $\tS$, $\tP$ & \\ [0.3em]
\toprule
& \multicolumn{2}{c}{\emph{Non-perturbative contributions (LO)}}\\
\midrule
\parbox[c]{1cm}{\centering LO\\ \centering $(Q/\Lambdabreak)^0$} &  \scalebox{0.8}{\CLOdiagram \OPEdiagram  \underbar{\Cdiagram{rectangle}{orange}{5pt}} } & - \\
\midrule
& \multicolumn{2}{c}{\emph{Perturbative contributions}}\\
\midrule
\parbox[c]{1cm}{\centering NLO\\ \centering $(Q/\Lambdabreak)^1$} & \scalebox{0.8}{\underbar{\Cdiagram{rectangle}{orange}{5pt}}}  & \scalebox{0.8}{\OPEdiagram} \\
\midrule
\parbox[c]{1cm}{\centering \NNLO\\ \centering $(Q/\Lambdabreak)^2$} & \scalebox{0.8}{\Cdiagram{rectangle}{orange}{5pt} \TPEbox \TPEloop \TPEtrianglel}  & - \\
& \scalebox{0.8}{\TPEcrossedbox  \TPEtriangler \underbar{\Cdiagram{diamond}{red}{7pt}}} & \\
\midrule
\parbox[c]{1cm}{\centering \NNNLO\\ \centering $(Q/\Lambdabreak)^3$} & \scalebox{0.8}{\TPEtrianglerr \TPEtrianglell \TPEloopl \TPEloopr}  & \scalebox{0.8}{\TPEbox \TPEloop \TPEtrianglel} \\
& \scalebox{0.8}{\underbar{\Cdiagram{diamond}{red}{7pt}} \underbar{\Ctriangle}} &\scalebox{0.8}{\TPEcrossedbox \TPEtriangler }  \\
\bottomrule
\end{tabular}
}
\label{tab:pot_diagrams}
\end{table}

{\it Formalism for perturbative bound-state calculations.$-$}
For \H, we implement Rayleigh-Schrödinger perturbation theory in J-NCSM with the Hamiltonian expanded in a spherical harmonic-oscillator basis of frequency $\omega$, including basis states up to $\Nmax$ oscillator quanta above the lowest configuration. At LO, we solve the Schrödinger equation 
\begin{equation}
    H^{(0)} \ket{\Psi^{(0)}_n} = E^{(0)}_n \ket{\Psi^{(0)}_n}, 
\end{equation}
non-perturbatively as described in Ref.~\cite{Thim:2025vhe}, to obtain the LO spectrum $\{E^{(0)}_n,\ket{\Psi^{(0)}_n}\}_n$. The LO Hamiltonian $ H^{(0)} = T + V^{(0)}$ is the sum of the intrinsic kinetic energy and the LO potential $V^{(0)}$. 

Perturbative corrections to the ground-state energy ($n=0$) are computed by insertions of the subleading potentials $V^{(\nu>0)}$ as 
\begin{align}
    E^{(1)} &= \braket{\Psi_0 |V^{(1)}|\Psi_0} \label{eq:E1_RS}\\
    E^{(2)} &= \braket{\Psi_0|V^{(2)}|\Psi_0} + \sum_{k\neq 0} \frac{|\braket{\Psi_0 | V^{(1)}|\Psi_k}|^2}{E^{(0)}_0-E^{(0)}_k} \label{eq:E2_RS}\\
    E^{(3)} &= \braket{\Psi_0|V^{(3)}|\Psi_0}  \nonumber\\
    &+2\sum_{k\neq 0} \frac{\braket{\Psi_0 |V^{(2)}|\Psi_k}\braket{\Psi_k |V^{(1)}|\Psi_0}}{E^{(0)}_0-E^{(0)}_k}\nonumber\\
    &+\sum_{k\neq 0}\sum_{m\neq 0} \frac{\braket{\Psi_0 | V^{(1)}|\Psi_k}\braket{\Psi_k | V^{(1)}|\Psi_m}\braket{\Psi_m |V^{(1)}|\Psi_0}}{(E^{(0)}_0-E^{(0)}_k)(E^{(0)}_0-E^{(0)}_m)} \nonumber\\
    &-\braket{\Psi_0 |V^{(1)}|\Psi_0} \sum_{k\neq0}\frac{|\braket{\Psi_0 | V^{(1)}|\Psi_k}|^2}{\left(E^{(0)}_0-E^{(0)}_k\right)^2},\label{eq:E3_RS}
\end{align}
where $\ket{\Psi_n} \equiv \ket{\Psi^{(0)}_n}$.
These sums are explicitly evaluated using the full LO spectrum. We find that the energy denominators do not suppress highly excited states enough to be neglected, and all intermediate states must be considered.

While it is feasible to compute the full LO spectrum in a J-NCSM calculation of \H, as done here, it is in general impossible to obtain more than a few converged eigenstates in many-body methods applied beyond the lightest nuclei. 
Indeed, when calculating \He{} and \Li{} with the M-NCSM method, we are restricted to iterative solutions for selected states, e.g., the ground-state, using the Lanczos algorithm. 

The corrections $E^{(\nu>0)}$ to the LO ground-state energy $E^{(0)}$, listed in \cref{eq:E1_RS,eq:E2_RS,eq:E3_RS}, can be obtained without explicit knowledge of the full spectrum by computing numerical derivatives of the selected eigenvalues to the Schrödinger equation. We add the subleading potentials multiplied by parameters $\bm{x}=(x_1,x_2,x_3) \in \mathbb{R}^3$ to the LO Hamiltonian and define
\begin{equation}
    H(\bm{x}) \equiv H^{(0)} + \sum_{\nu=1}^3 x_\nu V^{(\nu)},
\end{equation}
for which the Schrödinger equation 
\begin{equation}
    H(\bm{x}) \ket{\Psi_n(\bm{x})} = E_n(\bm{x}) \ket{\Psi_n(\bm{x})},
\end{equation}
can be solved, yielding $\bm{x}$-dependent states and eigenvalues.
The perturbative corrections to ground-state energy can be formulated as derivatives of $E_0(\bm{x})$ as
\begin{align}
    E^{(1)} &= \partial_{x_1} E_0(\bm{x}),  \label{eq:E1_FD}\\ 
    E^{(2)} &= \partial_{x_2} E_0(\bm{x}) + \frac{1}{2} \partial^2_{x_1} E_0(\bm{x}),  \label{eq:E2_FD} \\ 
    E^{(3)} &= \partial_{x_3} E_0(\bm{x}) + \partial_{x_1}\partial_{x_2} E_0(\bm{x}) + \frac{1}{6} \partial^3_{x_1} E_0(\bm{x}), \label{eq:E3_FD}
\end{align}
where all derivatives are evaluated at $\bm{x}=0$. This approach was applied at NLO in Refs.~\cite{Yang:2020pgi,Mondal:2025wml} and we will refer to it as the finite-difference (FD) method.

We perform numerical derivatives with finite-difference stencils \cite{Fornberg:1988fd}, where sufficient numerical accuracy is needed to compute third-order derivatives. If the eigenvalue, $E_0(\bm{x})$, is computed with precision $\epsilon$ (e.g.\@ $\epsilon=10^{-7}$ for single precision) the relative error in an $n$:th order derivative follows the general form $E_\epsilon(h) = \epsilon h^{-n} + K h^p$, where $h$ is the stencil step length. The first term is the floating point round-off error and the second is the error from using a finite-order ($p\in \mathbb{N}$) stencil, where $K$ is a constant. Thus, lowering $h$ will lower the stencil truncation error but increase the numerical round-off error---especially for the higher-order derivatives. A complete computation up to \NNNLO{} with $h^2$ ($h^4$) stencils requires 13 (31) exact diagonalizations for different values of $\bm{x}$.

\begin{figure}[t]
	\centering 
	\includegraphics[width=\columnwidth]{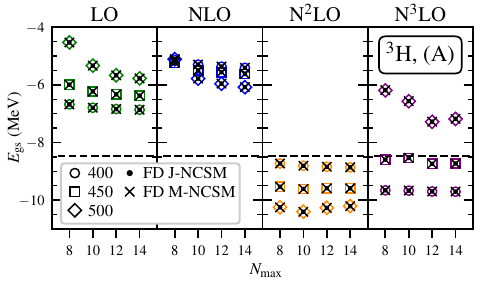}	
	\caption{Ground-state energy of \H{} as a function of $\Nmax$ at LO (leftmost panel) to \NNNLO{} (rightmost panel). The circles, squares, and diamonds show the exact Rayleigh-Schrödinger result for regulator cutoffs $\Lambda=400,450,500$~MeV, for interaction (A). The black dots (crosses) show the FD computations with the J- (M-) NCSM codes, respectively for $h=0.06$ and $p=2$. The dashed horizontal line shows the experimental \H{} ground-state energy \cite{Huang:2021nwk}. The HO frequency $\omega=24$~MeV is employed.} 
	\label{fig:H_conv}
\end{figure}

In \cref{fig:H_conv} we show the ground-state energy predictions for \H{} at orders N$^\nu$LO, i.e. $\sum_{k=0}^{\nu} E^{(k)}$, using the interaction from Ref.~\cite{Thim:2024yks} (referred to as interaction (A)) and compare the exact Rayleigh-Schrödinger computations to the results from the FD method. This comparison demonstrates that the FD method can achieve excellent accuracy despite working in a finite basis and at finite numerical precision. We also note that the accuracy is not dependent on $\Nmax$, which is important since only relatively low values of $\Nmax$ can be accessed in the \He{} and \Li{} computations. 

For \He{} and \Li{}, we cannot compare the FD method to exact Rayleigh-Schrödinger computations. We instead perform an extended convergence analysis, presented in the End Matter, applying stencils with different truncation orders to confirm convergence. We find that single precision ($\epsilon = 10^{-7}$) in the ground-state eigenvalue obtained from Lanczos diagonalization is sufficient for obtaining sub-percent numerical error in the predicted ground-state energy of \He{} up to \NNNLO, employing a $h^2$ stencil with $h=0.06$.

Finally, note that the energy correction at order $\nu$ is linear in the LECs at that same order, since they appear only in the first expectation value in \cref{eq:E1_RS,eq:E2_RS,eq:E3_RS}. This simplifies LEC inference using the \H{} ground-state energy, since a linear decomposition enables efficient evaluation across different LEC sets.

{\it Predicted ground-state energies of $^4$He and $^6$Li \label{sec:pred}.$-$}

\begin{figure}[t]
	\centering 
	\includegraphics[width=\columnwidth]{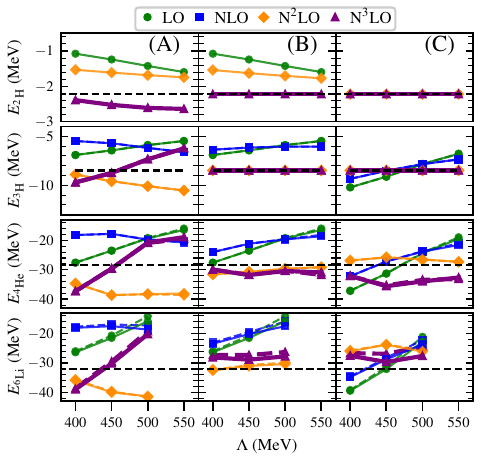}	
	\caption{Ground-state energies for $^{2,3}$H, \He{} and \Li{} (rows) as a function of cutoff for three interactions (columns) from \LO{} to \NNNLO. Interaction (A) is calibrated using neutron-proton phase shifts alone while (B) and (C) additionally use few-body input: \H{} at N$^{2,3}$LO and $^{2}$H at \NNNLO{} (B) or $^{2}$H already at \LO{} and \H{} at N$^{2,3}$LO (C). The solid (dashed) lines show $\Nmax=30 \ (28)$ for \H, $\Nmax=16 \ (14)$ for \He, and $\Nmax=14 \ (12)$ for \Li. The black dashed lines show the experimental values. The HO frequency $\omega=24$~MeV in all computations.} 
	\label{fig:2H_to_6Li_all_int}
\end{figure}
The \H{} result presented in the previous section, together with the convergence study, shows that the FD computations for $A>3$ are reliable. We now move on to compute ground-state energies in $^{2,3}$H, \He{} and \Li{} using three different interactions, (A) -- (C), with two additional variations discussed in the End Matter. 
The differences between these interactions are limited to the $^1S_0$ and $\tS$ channels. For higher partial waves, all interactions are identical and are calibrated to neutron-proton phase shifts following the procedure in Ref.~\cite{Thim:2024yks}.

We begin by considering interaction (A), taken from Ref.~\cite{Thim:2024yks}, where the LECs are constrained solely by neutron-proton phase shifts.  For this interaction, the ground-state energies of light nuclei, predicted up to \NNNLO{}, are poorly reproduced, see the left column of \cref{fig:2H_to_6Li_all_int}. Moreover, unnaturally large sub-leading contributions and a pronounced cutoff dependence are observed, particularly at \NNNLO{}.

We then explore how predictions for \He{} and \Li{} are impacted by gradually conditioning the inference of the $S$-wave LECs on the binding energies of $^{2,3}$H. This study is prompted by the observation that $A>2$ energies are very sensitive to the LECs in the $\tS$ channel, and in particular to the LEC $D^{(\nu)}_{SD}$ controlling the strength of the tensor interaction, and thus the mixing angle $(\epsilon_1)$. For interaction (B), the LECs in the $\tS$ channel are calibrated to reproduce the \H{} ground-state energy beyond NLO (where EFT truncation errors are expected to be smaller and the risk of overfitting is correspondingly reduced). Additionally, the deuteron binding energy is included at \NNNLO. 
The two NLO LECs in the $^1S_0$ channel are also fitted slightly differently than (A) to decrease the magnitude of the correction in this channel.
These choices lead to a substantial improvement in the predicted ground-state energies of \He{} and \Li, with the associated cutoff dependence markedly reduced, as shown in the middle column of \cref{fig:2H_to_6Li_all_int}.
Calibration details are provided in the End Matter together with a discussion of the interaction variant (B') that does not include the deuteron data in the fit.

A common finding for interactions (A) and (B) is that $^{2,3}$H are noticeably underbound at LO.  To enhance attraction in $S$-wave channels, we introduce a variation of interaction (A) in which the deuteron binding energy is included already at LO. Motivated by (B), we also include \H{} beyond \NLO{} and tune the calibration to reduce the repulsive strength of the $^1S_0$ NLO correction. The results of this interaction, labeled (C), are shown in the third column of \cref{fig:2H_to_6Li_all_int}. We find that \H{} becomes more bound at \NLO{} while predictions for \He{} and \Li{} are similarly improved as for (B) with the notable exception that the \NNNLO{} prediction for \He{} is considerably overbound.

From the results shown in \cref{fig:2H_to_6Li_all_int}, we conclude that accurately reproducing the \H{} ground-state energy at \NNLO{} and \NNNLO{} is essential for accurate predictions in \He{} and \Li. We further observe slightly better order-by-order convergence in interaction (B) than in (C), indicating that a stronger binding at LO is not necessarily advantageous.

Neutron-proton phase shifts and deuteron properties for the interactions shown in \cref{fig:2H_to_6Li_all_int} are presented in the Supplemental Material \cite{sup-mat}. An important observation is that the reduced cutoff dependence observed in \cref{fig:2H_to_6Li_all_int} at \NNNLO{} for interactions (B) and (C) translates into an increased cutoff dependence in the $\tS$ mixing angle. This likely results from our particular choice of calibration scheme, reinforced by the absence of a three-nucleon force that should enter at this order (or possibly even at \NNLO{} \cite{Friar:1996zw}). It is also possible that the exceptional cutoff effects observed in Ref.~\cite{Thim:2025vhe} impact the cutoff dependence already at the relatively small cutoffs explored here.

We note that \Li{} is at or above the $^4\mathrm{He}  +  {}^2\mathrm{H}$ threshold with several of the interactions, although model-space convergence in \Li{} needs to be analyzed more carefully~\cite{Forssen:2017wei}. This finding is consistent with observations in Ref.~\cite{Yang:2020pgi} and also seen at LO in WPC \cite{LENPIC:2018lzt}. This situation calls for further investigation, particularly related to the calibration of the NN force in the $P$-wave channels.

\begin{figure}
	\centering 
	\includegraphics[width=\columnwidth]{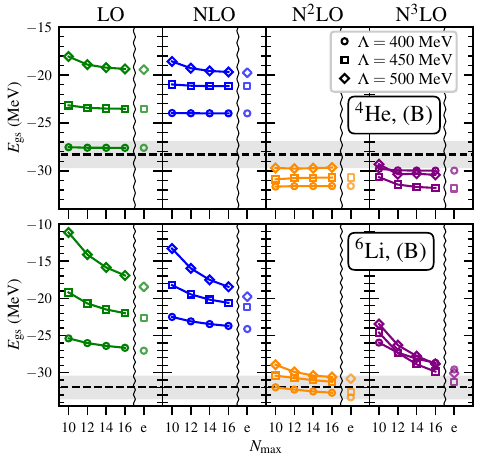}	
	\caption{Ground-state energies of \He{} (top) and \Li{} (bottom) as a function of $\Nmax$ at LO (leftmost panels) to \NNNLO{} (rightmost panels). Results for cutoffs $\Lambda=400, 450, 500$~MeV and interaction (B) are presented. The column denoted 'e' displays an exponential extrapolation in $\Nmax$ (where possible). The black dashed line shows the experimental ground-state energies \cite{Huang:2021nwk} and the shaded bands $\pm 5 \%$. We use $h=0.06$ with a $p=2$ stencil and $\omega=24$~MeV in all computations.} 
	\label{fig:He_Li_Nmax_conv}
\end{figure}

We conclude that interaction (B) provides the most accurate ground-state energies for \He{} and \Li. \Cref{fig:He_Li_Nmax_conv} shows the model-space convergence of the \He{} and \Li{} ground-state energies up to \NNNLO{} and $\Nmax=16$ computed with interaction (B). For \He{}, the energies are well converged through \NNNLO. Although convergence is somewhat slower in \Li, we still achieve a precision of $\approx 1$~MeV in the \NNNLO{} prediction. We also observe a marginally slower $\Nmax$ convergence with increasing order. These results demonstrate that converged perturbative computations are feasible using model spaces comparable to those required in non-perturbative computations. Assuming a breakdown scale near the nucleon-delta mass splitting \cite{Thim:2024yks}, a conservative estimate of the EFT truncation error at \NNNLO{} is $\approx 5 \%$, shown as the shaded band in \cref{fig:He_Li_Nmax_conv}. The observed chiral order-by-order convergence indicates that it is indeed possible to construct interactions in this PC that yield quantitatively accurate predictions for bulk observables in light nuclei.

{\it Summary and conclusions.$-$}
We have shown that accurate perturbative computations of the ground-state energies in \H, \He, and \Li{} up to \NNNLO{} can be obtained in the NCSM  using numerical derivatives from a set of Lanczos diagonalizations. We validated the numerical accuracy of this approach in \H{} by comparing with conventional Rayleigh-Schrödinger perturbation theory. We see no hurdles in applying this versatile numerical technique in other few- and many-body methods, also for higher mass numbers, since only non-perturbative solutions to the Schrödinger equation are needed.

Many-body computations are currently limited to relatively low values of the momentum cutoff, making it a challenge to analyze the RG invariance of predictions in all but selected few-nucleon Faddeev- and Faddeev-Yakubovsky calculations \cite{Song:2016ale,Konig:2019xxk}.

We have constructed and evaluated several different interactions with LECs calibrated using neutron-proton phase shifts and $^{2,3}$H ground-state energies. We found that different LEC calibration procedures had a large impact ($\approx10$~MeV) on predicted ground-state energies for \He{} and \Li. In particular, the predictions for the ground-state energies of $A>3$ systems were significantly improved by tuning the LECs in the $\tS$ channel to reproduce the \H{} binding energy. This induced a cutoff dependence in the $\tS$ mixing angle at \NNNLO{}, which may be remedied by the inclusion of three-nucleon forces.

Future developments thus include accounting for three-nucleon forces, Coulomb interactions and isospin-breaking, and performing a multi-order Bayesian inference of the LECs with \chEFT{} uncertainties. Enabling technologies like automatic differentiation \cite{Carlsson:2015vda} and emulators \cite{Frame:2017fah,Konig:2019adq,Duguet:2023wuh} tailored to perturbative computations further pave the way for systematically improvable \emph{ab initio} calculations rooted in quantum chromodynamics.

\begin{acknowledgments}
This work was supported by the Swedish Research Council (Grants No.~2020-05127, No.~2021-04507, No.~2024-04681, and No.~2025-05618). The computations were enabled by resources provided by Chalmers e-Commons at Chalmers and data handling by resources provided by the National Academic Infrastructure for Supercomputing in Sweden (NAISS), partially funded by the Swedish Research Council through grant agreement no. 2022-06725.
\end{acknowledgments}

\bibliography{bib}
\clearpage

\section{End Matter}

{\it Numerical accuracy of perturbative computations.$-$}
We now analyze the convergence of the FD method for computing perturbative corrections in more detail. The relative error in an $n$:th-order finite-difference derivative follows from $E_\epsilon(h) = \epsilon h^{-n} + K h^p$, where $\epsilon$ is the numerical precision of the function being differentiated, and $K h^p$ is the stencil truncation error. Optimal accuracy is obtained for sufficiently small $h$, prior to the onset of round-off error dominance.

\begin{figure}
	\centering 
	\includegraphics[width=\columnwidth]{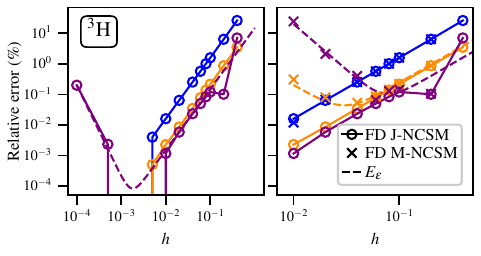}	
	\caption{Relative error in FD computations of $E^{(\nu)}$ with J-NCSM (circles) and M-NCSM (crosses) compared to the exact Rayleigh-Schrödinger computations for NLO (blue), \NNLO{} (orange), and \NNNLO{} (purple). The dashed line shows the error model $E_\epsilon(h)$ for $\epsilon=2\times 10^{-15}$ ($\epsilon=2\times 10^{-7}$) in the left (right) panel. All computations use $\Nmax=14$, $\Lambda=450$~MeV, $\omega=24$~MeV, and a $p=2$ stencil.} 
	\label{fig:H3_conv_h}
\end{figure}

We compare Rayleigh-Schrödinger and FD computations for \H{} in \cref{fig:H3_conv_h}. The left panel shows the relative error between the two methods for J-NCSM, and the dashed line displays the error model, $E_\epsilon(h)$, for $p=2$ and $\epsilon=2\times10^{-15}$, compared to the \NNNLO{} result. This demonstrates that the observed error in the FD computations follows the expected form consistent with double numerical precision, and that a relative error of $10^{-4} \ \%$ is achievable for an optimal choice of $h$. The right panel of \cref{fig:H3_conv_h} shows the relative error for both J-NCSM and M-NCSM, and the dashed line shows $E_\epsilon(h)$ for $\epsilon=2\times 10^{-7}$, consistent with single precision. This clearly demonstrates that the output precision is lower for the M-NCSM computations; however, a sub-percent error in all derivatives is still achievable for $h=0.06$.

\begin{figure}
	\centering 
	\includegraphics[width=\columnwidth]{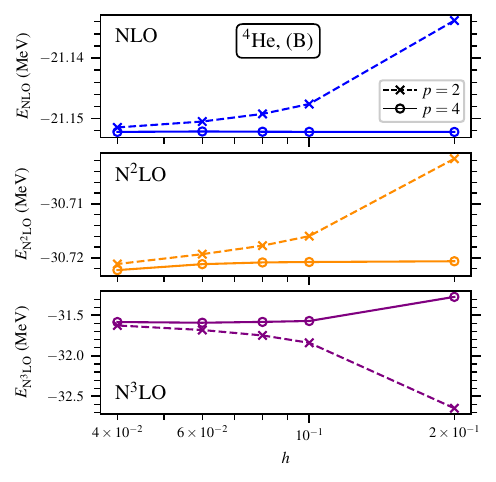}	
	\caption{Ground-state energy of \He{} at NLO (top panel) to \NNNLO{} (bottom panel) computed perturbatively with M-NCSM using the FD method. Results are shown for stencils with $p=2$ (dashed lines) and $p=4$ (solid lines). All computations use $\Nmax=14$, $\Lambda=450$~MeV, and $\omega=24$~MeV.} 
	\label{fig:He4_conv_stencil}
\end{figure}

For the \He{} and \Li{} computations, we do not have exact solutions to compare with. From the analysis in \H{} we know that the minimum error in the \NNNLO{} computation likely is achieved for $0.06 \leq h \leq 0.1$, and we perform computations in this interval. A converged result in this $h$-interval is confirmed by applying a higher-order stencil, which will lower the error from the term $K h^p$. 
\Cref{fig:He4_conv_stencil} shows the \He{} ground-state energy for stencils with $p\in \{2,4\}$. It is observed that the higher-order stencil enjoys a faster convergence for larger values of $h$, and for $0.06 \leq h \leq 0.08$ we observe a sub-percent error between the two stencils. Based on these results, we continue with computations in \He{} and \Li{} focusing on $p=2$ and $h=0.06,0.08$.

{\it Calibration of interactions.$-$}
We employ five different interactions in this work, with (A)--(C) shown in the main text and two variants (B'), (C') introduced here.
For these interactions, the LECs (shown in \cref{tab:LECs}) are calibrated using different sets of data. For high partial waves, the interactions are identical and calibrated according to the procedure in Ref.~\cite{Thim:2024yks} from the phase shift data listed in \cref{tab:LECs_fit_points_P}.
In the $^1S_0$ and $\tS$ channels, we explore five different ways of calibrating the LECs using the calibration data shown in \cref{tab:LECs_fit_points}. The $S$-wave phase shifts for all interactions are included in the Supplemental Material \cite{sup-mat}.

Interaction (A) is calibrated as in Ref.~\cite{Thim:2024yks} using only neutron-proton scattering phase shifts. Interaction (B') is a modification of (A) where the mixing angle calibration datum is replaced by the \H{} binding energy at \NNLO{} and \NNNLO. We also apply a shift in the $\Tl=25$~MeV calibration phase shift in $^1S_0$, to reduce the repulsive strength of the NLO correction. 
Note that this shift is corrected at subsequent orders, while we achieve mitigation of large alternating corrections. Interaction (B) is calibrated identically to interaction (B'), with the exception that the deuteron binding energy is added as calibration data at \NNNLO{} in interaction (B), as seen in \cref{tab:LECs_fit_points}.

In interaction (C'), we change the LO calibration data from phase shifts to the singlet scattering length $(a_s)$ and the deuteron ground-state energy $(E_d)$. Otherwise, the calibration follows that of interaction (A). Interaction (C) is a variation of (C') where the mixing angle calibration datum is replaced by the \H{} binding energy at \NNLO{} and \NNNLO. The shift in the $\Tl=25$~MeV calibration phase shift in $^1S_0$, is also introduced in (C).

\begin{figure}
	\centering 
	\includegraphics[width=0.75\columnwidth]{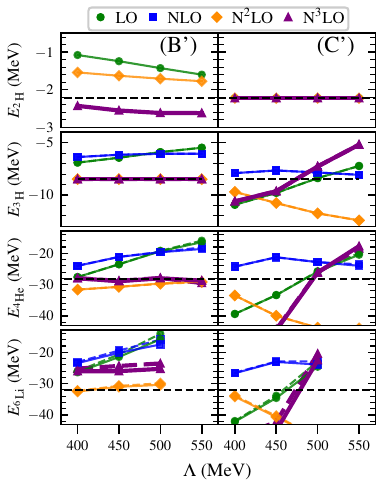}	
	\caption{Ground-state energies for $^{2,3}$H, \He{} and \Li{} (rows) as a function of cutoff for interactions (B') and (C'). The former is calibrated with neutron-proton phase shifts and \H{} at N$^{2,3}$LO. Interaction (C') is fitted differently for $S$-waves employing the $^{2}$H energy for all orders together with the singlet scattering length at \LO{}. The solid (dashed) lines show $\Nmax=30 \ (28)$ for \H, $\Nmax=16 \ (14)$ for \He, and $\Nmax=14 \ (12)$ for \Li. The black dashed lines show the experimental values. The HO frequency $\omega=24$~MeV in all computations.} 
	\label{fig:prim_small2}
\end{figure}

{\it Predicted energies for interactions (B') and (C').$-$}
The predicted ground-state energies in $^{2,3}$H, \He{} and \Li{} for interactions (A) -- (C) are shown in \cref{fig:2H_to_6Li_all_int}, while the results for interactions (B') and (C') are shown in \cref{fig:prim_small2}. Interaction (B') displays a noticeable underbinding for all systems up to NLO, and shows a similar description of \He{} and \Li{} as (B). The enhanced $S$-wave attraction in (C') and (C) addresses the underbinding at the lower orders. 
However, we observe that the predictions of interaction (C'), which do not include \H{} in the calibration, are quite poor.

\begin{table}
    \caption{LECs present in NN channels up to \NNNLO{}, see Ref.~\cite{Thim:2024yks} for details.}
\centering
\begingroup
\renewcommand{\arraystretch}{1.6} 
\begin{tabular}{c|c|c|c|c}
    Channel & LO & NLO & \NNLO & \NNNLO\\
    \hline\hline
    $^1S_0$ & $C^{(0)}_{^1S_0}$ & $C^{(1)}_{^1S_0}$, $D^{(0)}_{^1S_0}$ & $C^{(2)}_{^1S_0}$, $D^{(1)}_{^1S_0}$,  & $C^{(3)}_{^1S_0}$, $D^{(2)}_{^1S_0}$, \\ 
    & & & $E^{(0)}_{^1S_0}$ & $E^{(1)}_{^1S_0}$, $F^{(0)}_{^1S_0}$ \\
    
     $^3P_0$ &$D^{(0)}_{^3P_0}$ & -& $D^{(1)}_{^3P_0}$,$E^{(0)}_{^3P_0}$ & $D^{(2)}_{^3P_0}$,$E^{(1)}_{^3P_0}$\\
     
     $^1P_1$ &- & -& $D^{(0)}_{^1P_1}$ & $D^{(1)}_{^1P_1}$\\
     
    $^3P_1$ &- & -& $D^{(0)}_{^3P_1}$  & $D^{(1)}_{^3P_1}$\\
    
    $\tS$ &$C^{(0)}_{^3S_1}$ &-&$C^{(1)}_{^3S_1}$,$D^{(0)}_{^3S_1}$,& $C^{(2)}_{^3S_1}$,$D^{(1)}_{^3S_1}$,\\ 
   &&&$D^{(0)}_{SD}$&$D^{(1)}_{SD}$ \\
    $\tP$ &$D^{(0)}_{^3P_2}$ & -& $D^{(1)}_{^3P_2}$,$E^{(0)}_{^3P_2}$, & $D^{(2)}_{^3P_2}$,$E^{(1)}_{^3P_2}$, \\
    &&&$E^{(0)}_{PF}$& $E^{(1)}_{PF}$\\
    \hline
\end{tabular}
\endgroup
\label{tab:LECs}
\end{table}

\begin{table}
    \caption{Laboratory scattering energies $T_\mathrm{lab}$ (in MeV) of the Nijmegen phase shifts~\cite{Stoks:1993tb} used to calibrate the values of the LECs in all channels beyond $S$-waves.}
\centering

\begingroup
\renewcommand{\arraystretch}{1.1} 
\begin{tabular}{c|c|c|c|c}
    Channel & \LO & \NLO & \NNLO & \NNNLO \\
    \hline\hline
    $^3P_0$ & 25 &-& 25, 50 & 75, 100 \\
    $^1P_1$ & -&- &50&50 \\
    $^3P_1$ & -&-& 50& 50 \\
    $\tP$ &  $^3P_2: 30$ & - & $^3P_2: 30,50.$& $^3P_2: 30,50.$\\
    &  & & $\epsilon_2: 50$& $\epsilon_2: 50$\\
    \hline
\end{tabular}
\endgroup
\label{tab:LECs_fit_points_P}
\end{table}

\begin{table}
    \caption{Laboratory scattering energies $T_\mathrm{lab}$ (in MeV) of the Nijmegen phase shifts~\cite{Stoks:1993tb} used to calibrate the values of $S$-wave LECs at each chiral order. The spin-singlet scattering length $a_s = -23.74$~fm \cite{NavarroPerez:2014ovp}, as well as the deuteron and \H{} binding energies $E_{^2H} = -2.22$~MeV and $E_{^3H} = -8.48$~MeV \cite{Huang:2021nwk} used, as indicated. The $(+ \Delta)$ indicates that the Nijmegen phase shift at $T_\mathrm{lab}=25$~MeV is shifted $+7^\circ$ to reduce the strength of the NLO correction, as seen in the NLO phase shift predictions in \cref{fig:S_phases_all}.}
\centering

\begingroup
\renewcommand{\arraystretch}{1.1} 
\begin{tabular}{c|c|c|c|c|c}
    \toprule
    Pot. & Partial wave & \LO & \NLO & \NNLO & \NNNLO \\
    \hline\hline
        & $^1S_0$      & 5  & 5, 25 & 5, 25, 50 & 5, 25, 50, 75 \\
    A & $^3S_1$      & 30 & -     & 30, 50     & 30, 50\\
        & $\epsilon_1$ &    &       & 50        & 50 \\
    \midrule
         & $^1S_0$      & 5  & 5, 25 $(+\Delta)$ & 5, 25, 50 & 5, 25, 50, 75 \\
    B' & $^3S_1$      & 30 & -               & 30, 50     & 30, 50\\
         & $\epsilon_1$ &    &                 & $E_{^3H}$        & $E_{^3H}$ \\
    \midrule
         & $^1S_0$      & 5  & 5, 25 $(+\Delta)$ & 5, 25, 50 & 5, 25, 50, 75 \\
    B& $^3S_1$      & 30 & -  & 30, 50     & $E_{^2H}$, 50\\
         & $\epsilon_1$ &    &                 & $E_{^3H}$        & $E_{^3H}$ \\
    \midrule
        & $^1S_0$      & $a_s$  & 5, 25 & 5, 25, 50 & 5, 25, 50, 75 \\
    C'& $^3S_1$      & $E_{^2H}$ & -  & $E_{^2H}$, 50     & $E_{^2H}$, 50\\
        & $\epsilon_1$ &    &                 & 50        & 50 \\
    \midrule
         & $^1S_0$      & 5  & 5, 25 $(+\Delta)$ & 5, 25, 50 & 5, 25, 50, 75 \\
    C& $^3S_1$      & $E_{^2H}$ & -  & $E_{^2H}$, 50     & $E_{^2H}$, 50\\
         & $\epsilon_1$ &    &                 & $E_{^3H}$        & $E_{^3H}$ \\
    \midrule
    \bottomrule
\end{tabular}
\endgroup
\label{tab:LECs_fit_points}
\end{table}

\clearpage
\onecolumngrid
\setcounter{page}{1}
\setcounter{equation}{0}
\setcounter{figure}{0}
\setcounter{table}{0}
\renewcommand{\theequation}{S\arabic{equation}}
\renewcommand{\thefigure}{S\arabic{figure}}
\renewcommand{\thetable}{S\arabic{table}}

\section*{Supplemental Material} 

\subsection*{$S$-wave phase shifts}
\begin{figure*}
	\centering 
	\includegraphics[width=\textwidth]{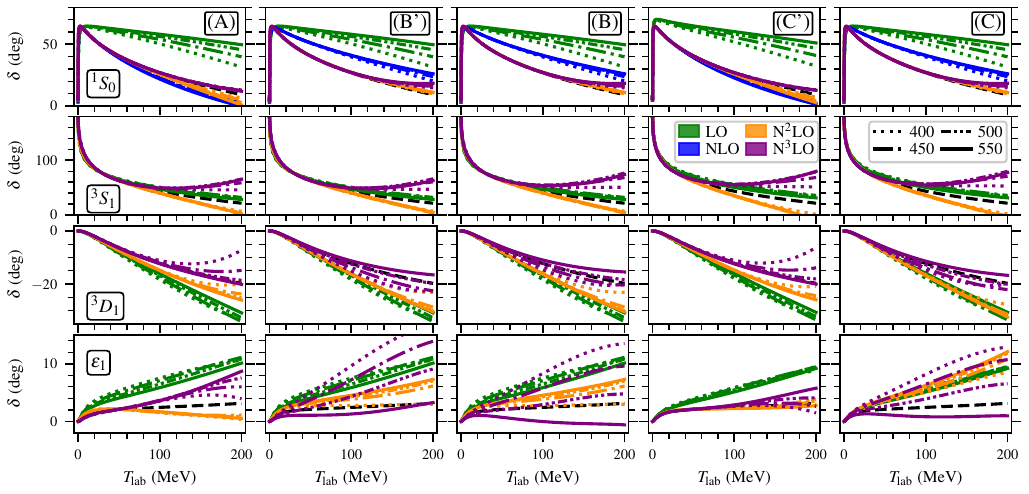}	
	\caption{Phase shifts in the $^1S_0$ and $\tS$ channels as a function of laboratory scattering energy, $T_\mathrm{lab}$ for various potentials (columns). Dotted, double-dotted-dashed, dot-dashed, and solid lines denote cutoffs $\Lambda=400,450,500,500$, respectively. The black dashed lines show the phase shifts from the Nijmegen partial wave analysis \cite{Stoks:1993tb}.} 
	\label{fig:S_phases_all}
\end{figure*}

The phase shifts in the $^1S_0$ and $\tS$ channels for all five interactions are shown in \cref{fig:S_phases_all}. Note that the NLO correction in the $^1S_0$ channel is large for (A) and (C'), which propagates to large alternating corrections in \H{} from the $^1S_0$ channel in $V^{(1)}$. This is mitigated in the (B'), (B), and (C) interactions via the $\Delta$ shift. An enhanced cutoff dependence can be observed in the mixing angle at \NNNLO{} for all interactions in which the \H{} binding energy is included.

\subsection*{Deuteron properties}
\begin{figure*}
	\centering 
	\includegraphics[width=\textwidth]{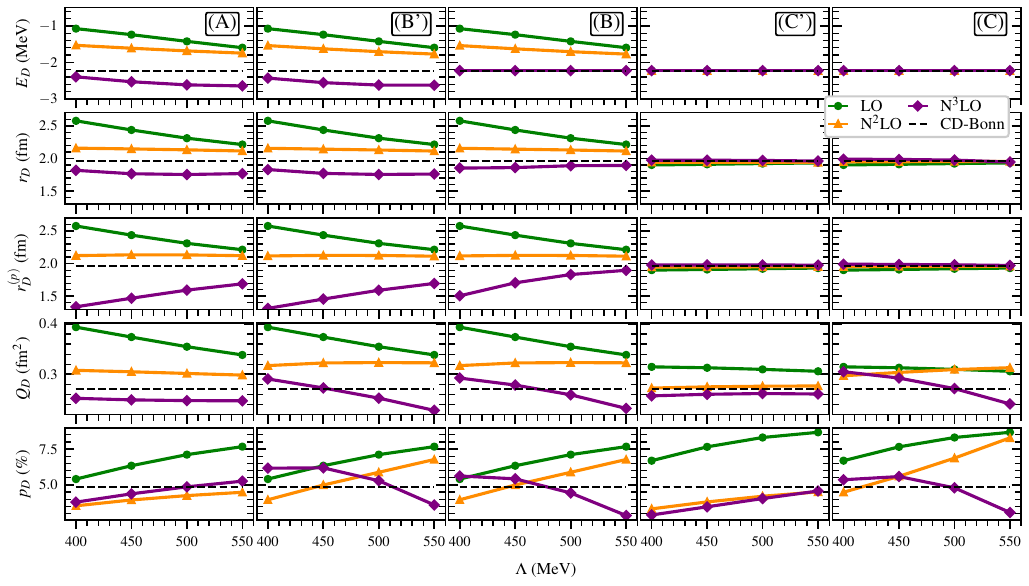}	
	\caption{Deuteron properties at LO to \NNNLO{} for all five interactions as a function of cutoff. The black dashed line shows the values from the CD-Bonn potential \cite{Machleidt:2000ge}. The row shows the deuteron binding energy, $E_D$. The following two rows show the deuteron radius computed in two different ways, as shown in \cref{eq:rD_exact,eq:rD_LO,eq:rD_N2LO,eq:rD_N3LO}. Finally, we show the quadrupole moment and the $D$-state probability in the deuteron wave function in the bottom two rows.} 
	\label{fig:deuteron_all}
\end{figure*}

We compute deuteron properties for all five interactions. The results are shown in \cref{fig:deuteron_all}. The deuteron wave functions are computed by solving the Schrödinger equation projected to the $\tS$ channel
\begin{equation}
    (T+V^{(0)})\ket{\psi^{(0)}_n} = E^{(0)}_n \ket{\psi^{(0)}_n},
\end{equation}
and sub-leading corrections are obtained as
\begin{equation}
    \ket{\psi^{(\nu)}_n} = \sum_{m\neq n} \frac{\braket{\psi^{(0)}_m| V^{(\nu)}|\psi^{(0)}_n}}{E^{(0)}_n-E^{(0)}_m} \ket{\psi^{(0)}_m}, \quad \nu=2,3.
\end{equation}
The expressions simplify since the NLO potential $V^{(1)}$ is zero in the $\tS$ channel. Furthermore, we define the full ground state wave function at order $\nu$ as
\begin{equation}
    \ket{\Psi^{(\nu)}_0} = \sum_{i=0}^v\ket{\psi^{(i)}_0}.
\end{equation}

There are two approaches to computing the deuteron properties, and observables in general. You can employ the perturbatively computed wave function $\ket{\Psi^{(\nu)}_0}$ as if it was a non-perturbative solution to the Schrödinger equation. In this case, the deuteron matter radius reads
\begin{equation}
    r^2_D = \braket{\Psi^{(\nu)}_0 | \hat{r}^2 | \Psi^{(\nu)}_0} = \sum_{i,j=0}^\nu \braket{\psi^{(i)}_0| \hat{r}^2 |\psi^{(j)}_0}
    \label{eq:rD_exact}
\end{equation}
One can also perform a strictly perturbative expansion, and only consider the terms in \cref{eq:rD_exact} up to the given order 
\begin{align}
    r^2_D &= \braket{\psi^{(0)}_0| \hat{r}^2 |\psi^{(0)}_0} + \left[\braket{\psi^{(2)}_0 | \hat{r}^2 | \psi^{(0)}_0} + \text{ h.c.}\right] \nonumber \\ &+ \left[\braket{\psi^{(3)}_0 | \hat{r}^2 | \psi^{(0)}_0} + \text{ h.c.}\right] + \dots
    \label{eq:rD_expanded}
\end{align}
where we define $\braket{\hat{r}^2}_{0} \equiv \braket{\psi^{(0)}_0 | \hat{r}^2 | \psi^{(0)}_0}$, and $\braket{\hat{r}^2}_{\nu} \equiv [\braket{\psi^{(\nu)}_0 | \hat{r}^2 | \psi^{(0)}_0} + \text{h.c.}]$, for $\nu=2,3$. By expanding the square root, the radius at each order reads
\begin{align}
    \mathrm{LO:} \quad r_D &= \sqrt{\braket{\hat{r}^2}_0}, \label{eq:rD_LO}\\
    \mathrm{N}^2\mathrm{LO:} \quad r_D &= \sqrt{\braket{\hat{r}^2}_0}\left(1 + \frac{\braket{\hat{r}^2}_2}{2\braket{r^2}_0}\right),   \label{eq:rD_N2LO}\\
    \mathrm{N}^3\mathrm{LO:} \quad r_D &= \sqrt{\braket{\hat{r}^2}_0}\left(1 + \frac{\braket{\hat{r}^2}_2 +\braket{\hat{r}^2}_3 }{2\braket{\hat{r}^2}_0}\right). \label{eq:rD_N3LO}
\end{align}
The strictly perturbative computations are denoted $r^{(p)}_D$ in \cref{fig:deuteron_all}. 

We note a discrepancy between the radius at \NNNLO{} computed with the expressions in \cref{eq:rD_exact} and \cref{eq:rD_LO,eq:rD_N2LO,eq:rD_N3LO} for interactions (A), (B'), and (B), but a good agreement for interactions (C') and (C). This is likely a result of the poor reproduction of the deuteron binding energy and radius at LO in interaction (A), (B'), and (B). The better reproduction of the radius at \NNNLO{} in the non-perturbative computation from \cref{eq:rD_exact} can be related to the improved convergence enjoyed by the so-called $Z$-parameterization \cite{Phillips:1999hh}, where the included higher-order terms in \cref{eq:rD_exact} improved the tail-description of the wave function relevant for the radius observable.

We compute the deuteron quadrupole moment $(Q_D)$ and $D$-state probability $(p_D)$ using the non-perturbative method similar to \cref{eq:rD_exact}, see, e.g., \cite{Machleidt:2000ge}. The results are also shown in \cref{fig:deuteron_all}.

\twocolumngrid

\end{document}